\documentclass[useAMS,usenatbib]{mn2e}
\usepackage{graphicx}
\title[The Broad-band Noise Characteristics of Selected AXPs and SGRs]
{The Broad-band Noise Characteristics of Selected Anomalous X-Ray Pulsars and Soft Gamma-ray
Repeaters} 
\author[B. K\"ulebi and \,\c{S}. Balman]{B. K\"ulebi$^{1}$\thanks{E-mail:\,bkulebi@astroa.physics.metu.edu.tr  Current Adress: Astronomisches Rechen Institut, M\"onchhofstr. 12-14, Heidelberg, Germany } and \,\c{S}. Balman$^{1, 2}$\thanks{E-mail:\,solen@astroa.physics.metu.edu.tr}
\\
$^{1}$Department of Physics, Middle East Technical University,
In\"on\"u Bulvar{\i}, Ankara, Turkey\\
$^{2}$Astrophysics Missions Division, Research and Scientific Support 
Department of ESA, ESTEC,\\ 
~$^{}$Postbus 299 SCI-SA, Keplerlaan 1 2201 AZ 
Noorwijk ZH, The Netherlands}
\begin{document}


\pagerange{\pageref{firstpage}--\pageref{lastpage}} \pubyear{2009}

\maketitle

\label{firstpage}

\begin{abstract}
We present the broad-band noise structure of selected Anomalous X-Ray Pulsars (AXPs) and Soft 
Gamma Repeaters (SGRs) in the 2-60 keV energy band. We have analyzed Rossi X-ray Timing 
Explorer Proportional Counter Array archival light curves for four AXPs and one SGR. We 
detect that the persistent emission of these sources show band limited noise at low frequencies 
in the range 0.005-0.05 Hz varying from 2.5$\%$ to 70$\%$ integrated rms in times of 
prolonged quiescence and following outbursts. We discovered band-limited red noise in 1E 
2259+586 only for $\sim$2 years after its major 2002 outburst. The system shows no broad-band 
noise otherwise. Although this rise in noise in 1E 2259+586 occurred following an outburst 
which included a rotational glitch, the other glitching AXPs showed no obvious change in 
broad-band noise, thus it does not seem that this noise is correlated with glitches. 
The only source that showed significant variation in broad-band noise was 1E 1048.1-5937, 
where the noise gradually rose for 1.95 years at a rate of $\sim$3.6$\%$ per year. For this source 
the increases in broad-band noise was not correlated with the large increases in persistent 
and pulsed flux, or its two short SGR-like bursts. This rise in noise did commence after  a long burst, however given the sparsity of this event, and the possibility that similar bursts 
went unnoticed the trigger for the rise is noise in 1E 1048.1-5937 is not as clear as for 
1E 2259+586. The other three sources indicate a persistent band-limited noise at low levels 
in comparison.
\end{abstract}


\section{Introduction}

Anomalous X-Ray Pulsars (AXPs) and Soft Gamma Repeaters (SGRs) are part of a new class of neutron stars (NSs), whose emission mechanisms commute with neither accreting X-Ray Pulsars nor radio pulsars (PSRs). Their periods are clustered between 2 - 12 s and they are characterized by their rapid spin down ($\sim 10^{-13}- 10^{-11}~\rm{s~s}^{-1}$) and high quiescent luminosities 
($\sim10^{33} - 5\times10^{35}~\rm{erg~s}^{-1}$) that can not be explained by spin down mechanisms. 
They also show wide range of variability; including flux and pulse profile variations 
(Dib, Kaspi \& Gavriil 2007a; Tam et al. 2008), glitches (Dib, Kaspi \& Gavriil 2007b), 
bursts (Gavriil, Kaspi \& Woods 2002) and strong timing noise (Baykal et al. 2000; Kaspi, Gavriil \& 
Chakrabarty et al. 2001). Three SGRs are known to exhibit giant bursts (SGR 0526-66, SGR 1900+14, 
SGR 1806-20), which in two of these objects rapid quasi-periodic oscillations (QPOs) 
are detected after the onset of their flares (SGR 1900+14; Strohmayer \& Watts 2005, SGR 1806-20; 
Israel et al. 2005). 

SGRs and AXPs are widely accepted as being magnetars, neutron stars (NSs) with magnetic fields strengths  greater than the quantum critical level ($10^{14-15}$ Gauss), whose main source of energy is the  dissipation of these fields (Duncan \& Thompson 1992; Thompson \& Duncan 1995). The magnetar model for these objects explains the SGR bursts and the large flares, and it predicted the AXP bursts which occurred in 1E 1048.1-5937 (Gavriil, Kaspi \& Woods 2002) and 1E 2259+586 (Gavriil \& Kaspi 2002). Thompson \& Duncan (1995) argue that the bursts are due to crustal cracking as a result of magnetic diffusion, and that the large flares are due to major reconfigurations of the  global magnetic field (Thompson \& Duncan 1995). Magnetar strength magnetic fields can also account for  the rapid spin down of these objects and the QPOs observed during large flares (e.g, Israel et al. 2005; Strohmayer \& Watts 2005). The existence of a fossil disk, and accretion from it, has also been proposed to explain the persistent properties of these sources (van Paradijs et al. 1995; Chatterjee et al. 2000; Alpar 2001). However, this model  has difficulty explaining the giant flares, repeat bursts and the pulsed optical emission that exists in this class of objects.  

In this paper, we investigated the properties of the broad-band noise in selected AXPs and SGRs, and present plausible origins for it within the context of the above two models. For example, if  AXPs and SGRs have fall-back disks that they accrete from, as has been suggested by  Ertan et al. (2007), then we can compare this noise to that of other accreting sources. For instance, the broad-band noise in X-ray binaries spans a very broad range and  is due to variable X-ray flux from the accretion disk. In corona hosting X-ray binaries, a contribution of the noise in the high frequency range is due to photon delays in Compton up scattered radiation (van der Klis 2006 and references therein). This is important given that AXPs and SGRs show hard power law like tails in their energy spectra that are also most likely due to photons being up-scattered in energy.  In the context of the magnetar model, these hard-tails in AXPs and SGRs have been successfully modeled by resonant cyclotron scattering (Lyutikov \& Gavriil 2006; Rea et al. 2008). We argue that resonant cyclotron scattering  can also introduce broad band noise in the frequency range observed for these sources. Thus, in order to discriminate between the two models it becomes of great interest to study the broad-band noise in AXPs and SGRs. 

\begin{table}
\caption {Observation log of the light curves analyzed in this work.}
\begin{tabular}{llccl} 
\hline
Object  &Obs Id         & Ave. Count Rate & Burst$\dagger$ \\
& & c/s & \\
\hline

1E 1048.1-5937  &       40083-08        &       3.889   & no \\
1E 1048.1-5937  & 50082-04      & 3.771 & no \\
1E 1048.1-5937  & 60069-03      & 3.302 & yes \\ 
1E 1048.1-5937  & 70094-02      & 7.030 & no \\
1E 1048.1-5937  & 80098-02      & 5.667 & no \\
1E 1048.1-5937  & 90076-02      & 3.580 & yes \\
1E 1048.1-5937  & 91070-02      & 3.255 & no \\
1E 1048.1-5937  & 92006-02      & 3.457 & no \\
1E 2259+586     & 20145-01      & 1.799 & no \\
1E 2259+586     & 20146-01      & 2.269 & no \\
1E 2259+586     & 40082-01      & 1.752 & no \\
1E 2259+586     & 50082-01      &       1.167   & no \\
1E 2259+586     & 60069-01      &       0.4600 & no \\
1E 2259+586     & 70094-01      & 2.695 & yes \\ 
1E 2259+586     & 80098-01      & 1.765 & no \\
1E 2259+586     & 90076-01      & 1.673 & no \\
RXS J1708-40    & 40083-14      & 13.02 & no \\
RXS J1708-40    & 50082-07      &       12.10   & no    \\
RXS J1708-40    &       60069-07        &       12.37   & no \\
RXS J1708-40    &       60412-01        &       12.99 & no \\
RXS J1708-40    &       70094-04        & 13.58 & no \\
RXS J1708-40    &       80098-04        &       13.38 & no \\
RXS J1708-40    & 90076-04      & 13.38 & no \\
1E 1841-045     & 40083-11      & 9.521 & no \\
1E 1841-045     & 50082-05      & 8.835 & no \\
1E 1841-045     &       70094-03        &       9.937 & no \\
1E 1841-045     & 90076-03      & 9.013 & no \\ 
1E 1841-045     & 91070-03      & 9.006 & no \\ 
SGR 1806-20     & 50142-01      & 7.626 & yes \\ 
SGR 1806-20     & 50142-03      & 7.607 & yes \\
SGR 1806-20     & 60121-01      & 7.053 & yes \\
SGR 1806-20     & 70136-02      & 10.69 & yes \\
SGR 1806-20     & 90073-02      & 10.93 & yes \\
SGR 1806-20     & 90074-02      & 12.70 & yes \\
SGR 1806-20 & 91062-02  &       8.721 & yes \\
\hline
\multicolumn{5}{l}{\ $\dagger$ Observation dates have been correlated with the GCN} \\ 
\multicolumn{5}{l}{\ (http://gcn.gsfc.nasa.gov/gcn3\_archive.html) burst data }\\
\multicolumn{5}{l}{\ and relevant papers (see captions of Fig. 5).}\\
\end{tabular}
\end{table}

\section{Data and Observations}
To calculate the band limited noise of these objects we used the 2 - 60 keV archived  public Rossi X-ray Timing Explorer (RXTE) light curves. The RXTE archive has a total  of 10 sources that are either AXPs or SGRs. We used only appropriate sets of data, eliminating sources depending on epochs, exposures, burst dates and objects in the field of view (FOV). 

We recovered 1 known SGR and 4 known AXP observations with fields relatively free of other sources. The exceptions are 1E 1048.1-5937 with Eta Carinea and SGR 1806-20 with XTE J1810-197 in their FOV, which both do not effect the noise levels in our analysis. The variability of Eta Carinae occurs in a very 
different time scale ($\sim$85 days; Corcoran et al. 1997) with respect to 256 seconds used in this study. XTE J1810-197 is a transient source which was quiescent during the observations in our analysis 
(Ibrahim et al. 2004). Other fields include strong and variable sources. The data were obtained by the Proportional Counter Array (PCA; Jahoda et al. 1996) instrument aboard RXTE. We obtained 125 ms resolution background subtracted light curves from archived standard products (StdProds) of RXTE. These light curves are screened through standard procedures; namely source elevation greater than $10^\circ$, pointing offset smaller than $0.02^\circ$, also PCU turn-off times are subtracted through defining good time intervals.  Backgrounds are generated by pcabackest command of FTOOLS with the use of background models of the relevant epoch. These net lightcurves also include the barycentrically corrected time columns, which were used in our analyses\footnote{All of the procedures involved in order to produce StdProds are explained in http://heasarc.gsfc.nasa.gov/docs/xte/recipes/stdprod\_guide.html}. Only when necessary, particularly to check background effects, non-background subtracted light curves were created from original data using SEEXTRCT v4.2. The manipulation of the data was made with the FTOOLS v5.21 software. 

In order to measure the broad-band noise, we derived and averaged several power spectra for each source.
The power spectral densities (PSD) expressed were calculated in terms of the fractional rms 
amplitude squared following Miyamoto et al. (1991), and the expected white noise levels 
were subtracted hence leaving us with the rms fractional variability of the time series in units of 
($\rm{rms}/\rm{mean})^2$/Hz.
We used in total 658 light curves for our analysis (151 for 1E 2259+586, 136 for 1E 1048.1-5937, 
125 for SGR 1806-20, 82 for 1E 1841-045 and 153 for RXS J1708-40), 
each of which was segmented into 10--50 individual power spectra to construct average PSD in a given time line.

\begin{figure}
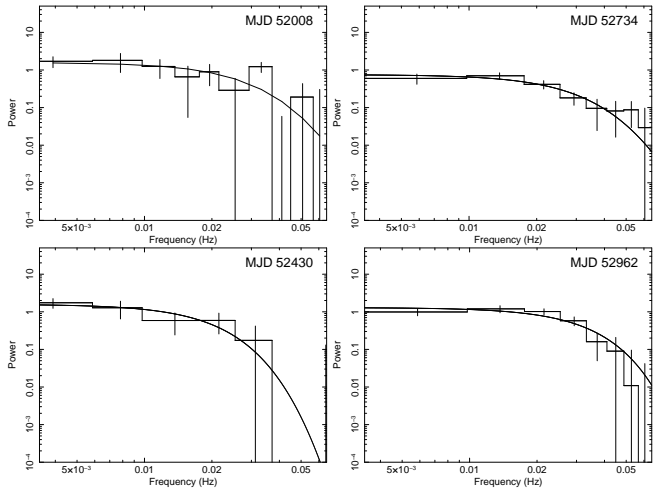

\centerline{
\includegraphics[scale=0.16,angle=270]{1048_60069_burstdahaonce_256_8_lb113_fit.ps} 
\includegraphics[scale=0.16,angle=270]{1048_80089_3ay1_fit.ps}}
\centerline{
\includegraphics[scale=0.16,angle=270]{1048_60069_sonra_fit.ps} 
\includegraphics[scale=0.16,angle=270]{1048_80089_3ay2_fit.ps}}
\caption{The averaged PSD in time showing the broad-band noise of AXP 1E 1048.1-5937. MJD 52211 and 52227 are burst 
epochs (Gavriil \&  Kaspi 2004). The time line increases top to bottom first, then from left to right.}
\end{figure}

\section{Analysis and Results}
In order to increase the signal to noise of the PSD to look for broad-band noise, the 
averaged light curves are grouped in three month intervals yielding 20 ks--50 ks long exposure times for each source. 
{\it The burst intervals noted in Table 1 are excluded from our data samples 
to avoid additional red noise because of the light curve variability during bursts. 
The remaining burst like features (especially for the bursting SGR 1806-20) and spurious events due to instrument errors are screened by hand from the light curves.} 
We averaged several 256 s long PSD for each source. 

Figures 1 through 4 show the representative averaged PSD fitted with a broken power law model; $\gamma\left[1.+ (x/\nu)^2 \right]^{-\alpha}$, in which $\nu$ is the frequency, $\gamma$ is the ($\rm{rms}/\rm{mean})^2$/Hz value of the flat part, $x$ is the cutoff frequency and $\alpha$ is the power law index. 
Each spectra in the analyses is averaged over three-month intervals and are separated by 5--900 days. In all sources, we find red noise with a break in the averaged PSD which may be considered band limited noise (BLN).

For all figures, the reduced chi squares of the fits are between 0.6-1.2, the range of observed break frequencies  are between 0.05-0.7 Hz and the power law indices are  between (-)0.5-375. 
Table 2 shows the list of power law indices derived from the fits to PSD of the sources in time. An irregular variation of the indices in time is apparent.  The ranges in Table 2 denote 1 $\sigma$ confidence level error ranges from fits. It is very hard to determine the value of these indices, since the decline in the power spectrum is near the edge of the sampling window and errors also increase considerably. We would like to make a note that for 2 $\sigma$ errors on the power law index, the canonical value of -2 is in the acceptable range. In addition, the fitted PSD are integrated over $5\times10^{-3}-5\times10^{-2}$ Hz interval to get the integrated fractional rms values. Time evolution of these values for 1E 1048.1-5937 and 1E 2259+586 are plotted in Figure 5.

We found that for 1E 1048.1-5937 and 1E 2259+586 (two AXPs), their burst (maybe also flare) 
epoch correlate with their broad-band noise levels. 
The steady $\sim15\%$ rms noise of 1E 1048.1-5937 showed a slight drop, during 
which two of its SGR-like short bursts and a flare appeared, to levels around $\sim10\%$ rms in 16 months. This low noise recovered throughout the next 5 months until it reached its former noise level (see Figures 1 and 5). The red noise level of the source rose after a long burst and later a glitch and large scale flux increase (Dib et al. 2007; Tam et al. 2008) to levels $\sim25\%$ rms and stayed at this level (with some variability) for about 1.95 years (see Fig. 5, top panel). 1E 2259+586 showed a slightly different behavior. It had no detectable broad-band red noise in quiescence and developed broad-band red noise around $\sim60\%$ 
rms level right after its burst active state and the on-set of a glitch, stayed consistently at this level throughout nearly 2 years, until it diminished in about 1.5 months (at MJD 53233) (see Figures 2 and 5 bottom panel) back to its original level of no broad-band  noise. 

The other two AXPs 1E 1841-045 and RXS J1708-40 maintain a persistent broad-band noise
level compared with the two AXPs discussed in the above paragraph.
Sources showed around $\sim11\%$ rms and  $\sim7\%$ rms level noise for 1E 1841-045 and 
RXS J1708-40, respectively (with slight increase in time). 
The total time span of the detected broad-band noise
for the two different epochs for RXS J1708-40 were 2.3 years (51215-52053 MJD) 
and 2.6 years (52366-53325 MJD) .
Also for 1E 1841-045 the time spans for two separate time epochs analized in this work
were   2 years (51977-51260 MJD)
and 3.5 years (53635-52349 MJD). 1E 1841-045 also stands out as the noisier of the two.
SGR 1806-20 exhibited broad-band noise 
slightly fluctuating about a persistent level of $\sim10\%$, overall. The time span
of our analysis in two epoch ranges were 1.4 years (51600-52100 MJD) and 1.7 years 
(53000-53600 MJD). We had difficulty relating the burst activity and the PSD for 
SGR 1806-20 due to the existence of excessive number of bursts. 


\begin{figure}
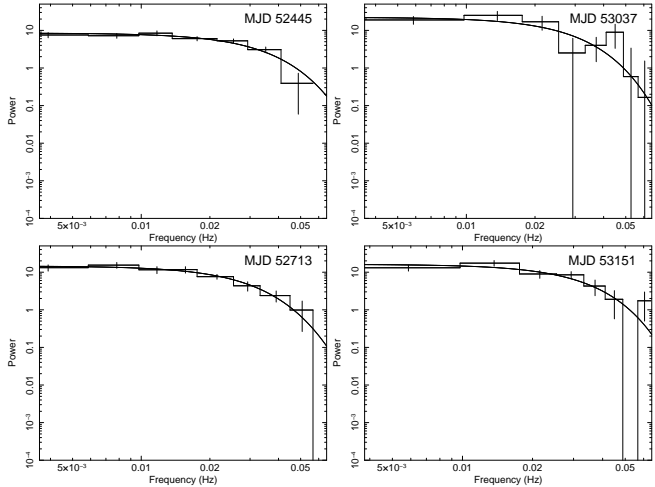

\centerline{
\includegraphics[scale=0.16,angle=270]{2259_70094_sonra_256_4_lb12_fit.ps} 
\includegraphics[scale=0.16,angle=270]{90076_3ay1_256_4_2_fit.ps}}
\centerline{
\includegraphics[scale=0.16,angle=270]{2259_80098_3ay2_256_4_lb115_fit.ps} 
\includegraphics[scale=0.16,angle=270]{90076_3ay2_256_4_2_fit.ps}}
 
\caption{The averaged PSD in time showing the broad-band noise in AXP 1E 2259+586 with MJD 52443 as 
the burst epoch (Kaspi et al. 2003). 
The time line increases top to bottom first, then from left to right.}
\end{figure}

\begin{table}
\caption {The power law indices of AXPs and SGRs derived from fitted PSD in this work.}
\begin{tabular}{lllll} 
\hline
Epoch (MJD)  &  Epoch range (d)  &  index(-) &   index error\\
\hline
{\bf 1E 1048.1-5937} &&& \\
51321   &       $\pm$   87      &       121.97  &       $\pm$   43.73   \\
52058   &       $\pm$   50      &       9.99    &       $\pm$   5.31    \\
52160   &       $\pm$   46      &       51.64   &       $\pm$   25.73   \\
52275   &       $\pm$   42      &       32.16   &       $\pm$   10.01   \\
52478   &       $\pm$   48      &       79.31   &       $\pm$   18.1    \\
52626   &       $\pm$   42      &       279.25  &       $\pm$   64.75   \\
52777   &       $\pm$   43      &       9.24    &       $\pm$   2.39    \\
53010   &       $\pm$   48      &       191.55  &       $\pm$   45.75   \\
\hline
{\bf 1E 2259+586} &&&\\
52490   &       $\pm$   45      &       29.07   &       $\pm$   4.5     \\
52601   &       $\pm$   53      &       4.77    &       $\pm$   2.17    \\
52757   &       $\pm$   44      &       90.01   &       $\pm$   38.79   \\
52863   &       $\pm$   50      &       17.94   &       $\pm$   3.35    \\
52975   &       $\pm$   47      &       256.15  &       $\pm$   65.75   \\
53087   &       $\pm$   50      &       64.87   &       $\pm$   27.39   \\
53192   &       $\pm$   41      &       144.65  &       $\pm$   34.05   \\
\hline
{\bf SGR 1806-20} &&&\\
51693   &       $\pm$   76      &       0.46    &       $\pm$   0.08    \\
51825   &       $\pm$   46      &       0.98    &       $\pm$   0.2     \\
52065   &       $\pm$   44      &       16.65   &       $\pm$   6.41    \\
53288   &       $\pm$   43      &       1.72    &       $\pm$   0.19    \\
53148   &       $\pm$   52      &       108.47  &       $\pm$   9.74    \\
53498   &       $\pm$   48      &       32.11   &       $\pm$   4.32    \\
\hline
{\bf 1E 1841-045} &&&\\
51334   &       $\pm$   74      &       18.37   &       $\pm$   5.29    \\
51534   &       $\pm$   63      &       6.32    &       $\pm$   2.06    \\
51788   &       $\pm$   60      &       24.13   &       $\pm$   9.54    \\
51931   &       $\pm$   46      &       78.41   &       $\pm$   38.69   \\
52415   &       $\pm$   66      &       30.36   &       $\pm$   5.78    \\
52598   &       $\pm$   69      &       30.79   &       $\pm$   7.52    \\
52803   &       $\pm$   77      &       124.95  &       $\pm$   33.76   \\
52940   &       $\pm$   40      &       45.38   &       $\pm$   10.03   \\
53125   &       $\pm$   52      &       39.15   &       $\pm$   9.16    \\
53486   &       $\pm$   46      &       211.9   &       $\pm$   57.5    \\
53590   &       $\pm$   44      &       13.37   &       $\pm$   2.47    \\
\hline
{\bf 1RXS J1708-40} &&&\\
51317   &       $\pm$   102     &       46.69   &       $\pm$   16.39   \\
51530   &       $\pm$   84      &       335.05  &       $\pm$   179.05  \\
51845   &       $\pm$   190     &       371.07  &       $\pm$   282.24  \\
52044   &       $\pm$   9       &       9.26    &       $\pm$   4.74    \\
52409   &       $\pm$   43      &       0.63    &       $\pm$   0.14    \\
52538   &       $\pm$   51      &       46.23   &       $\pm$   11.45   \\
52681   &       $\pm$   37      &       375.75  &       $\pm$   172.65  \\
52794   &       $\pm$   49      &       26.86   &       $\pm$   8.55    \\
52906   &       $\pm$   47      &       160.75  &       $\pm$   30.15   \\
53009   &       $\pm$   49      &       66      &       $\pm$   26.16   \\
53119   &       $\pm$   35      &       24.87   &       $\pm$   6.61    \\
53206   &       $\pm$   47      &       9.83    &       $\pm$   2.04    \\
53291   &       $\pm$   34      &       53.45   &       $\pm$   14.31   \\
\hline
\end{tabular}
\end{table}

\section{Discussion}

Figure 5, shows the comparative evolution of the band-limited red noise in 1E 1048.1-5937  and 1E 2259+586. The ordinate is the integrated rms band-limited noise and the abscissa is the time in MJD. The error in time (x error) is the total time of the accumulated observation IDs for each data point. The y error is the 1 $\sigma$ error in the integrated rms values.   
1E 2259+586 showed the highest level of noise and longest duration of change. 
1E 1048.1-5937  is the only source that showed a drop of the noise level assuming level at start of 
MJDs is the steady normal level of noise. It was also the most noisy source after 1E 2259+586. 
A significant rise of the rms broad-band noise level in 1E 1048.1-5937 (like 1E 2259+586) 
after the long-burst epoch at 53185 MJD is detected for about 1.95 years (may have lasted 
about 3 years judging from telegrams; Rea et al. 2007). 

On the basis of torque noise, 1E 2259+586 is very quiet (Baykal et al. 2000; Gavriil  \& Kaspi 2002),
but after its outburst and a glitch, 1E 2259+586 experienced rapid spin down over 60 days
(Woods et al. 2004) and showed an excess in infrared radiation (Hulleman et al. 2001). In this period 
1E 2259+586 displayed band-limited noise  lasting for a timescale of 2 years and no band-limited 
red noise other wise. 1E 1048.1-5937 shows a highly variable torque noise such that, it can not be 
phase connected for long time intervals which is consistent with accreting sources 
(Baykal et al. 2000; Kaspi, Gavriil \& Chakrabarty et al. 2001). In our work, this source showed 
band-limited red noise at all times. The substantial increase of the broad-band noise of 
1E 1048.1-5937 in Figure 5  
(from 15$\%$ to about 25-30$\%$) occured gradually after the the long burst ($>$ 699 sec) at 53185 MJD (Gavriil, Kaspi \& Woods 2006). Following this, there has been a flux increase (outburst-like) from 1E 1048.1-5937 reported along with a spin-up glitch at 54186 MJD (Dib et al. 2007; Tam et al. 2008). The decrease of this rise of the soft X-ray flux was reported to be at 54265 MJD (Rea et al. 2007; Tam et al. 2008). These latest dates are at the edge of our curve for 1E 1048.1-5937 in Figure 5 and are not included in our results.
However, it is clear that it follows our peak of integrated rms values for the source. {\it We note that the rise in broad-band noise started much earlier than the flux outburst right after the long burst at 53185 MJD (Gavriil, Kaspi \& Woods 2006).}
On the other hand, we caution that given the sparsity of this event, and the possibility that similar
bursts went unnoticed, this start for the rise in noise is not as clear as in 1E 2259+586. We also
note that this increase in red noise is not very correlated with a large increase in persistent and 
pulsed flux in the early stages until the outburst-like flux increase.
Prior to its burst at 53185 MJD and the broad-band noise increase, there was a slight decrease of red noise during which two SGR-like short bursts (52211 MJD and 52227 MJD) and two flares (52218 MJD and 52444 MJD) appeared (Gavriil, Kaspi \& Woods 2002; Gavriil \& Kaspi 2004). The two bursts showed 
characteristics similar to SGR bursts like marginal pulsed flux increase not more than 
3$\sigma$ and short duration ($<$50 s 1st burst and $<$ 2 s 2nd burst). Since we did not detect 
any correlation of broad-band noise with the SGR bursts in SGR 1806-20, we do not expect 
any effects here. The first flare starts between the two SGR-like bursts at 52218 MJD, 
after which an epoch of lower rms noise started. The drop of the broad-band noise showed a 
brief increase in the second flare (maximum at 52444.4 MJD, see Fig. 5) and the  trend of 
decrease and recovery of noise after the flare is similar to the torque variability that was 
observed by Tam et al. (2008).

The time-averaged PSD of the four AXPs and the SGR showed in general very low frequency noise. 
We detected that this noise to be band-limited and in the frequencies higher than 0.05 Hz. 
There were little or no noise at all times below these frequencies given the sensitivity of RXTE (also background) and statistical quality of RXTE light curves. 
The broad-band noise of AXPs and SGRs can be compared with the other isolated pulsars or X-ray binaries on the basis of their origin (e.g., accretion process, coronal activity, 
rotational variation). Broad-band noise in X-ray binaries is a result of
aperiodic X-ray flux variability in the form of continuum noise and QPOs with frequencies ranging from
several mHz to more than thousand Hz (Hasinger $\&$ van der Klis, 1989, van der Klis 2000,2006). 
The flattening in the lower frequencies resemble PSD in disk-fed X-ray binaries, specifically the low hard state (LS) noise of black hole candidates and the very low frequency noise (VLFN) of low-magnetic field neutron star LMXBs 
(van der Klis, 2006 and references therein). A flat topped noise arises in black hole candidates with 
large integrated rms noise level ($\sim50\%$) in the low frequencies (van der Klis 2006). 
The VLFN of LMXBs show the flat topped noise with a few percent integrated rms and the continuum 
power law component shows indices between (-)1.5-2, in general a range which holds for all the 
X-ray binaries (van der Klis 2006). 
The VLFN of the X-ray binaries is correlated  with $\dot{M}$ variations associated with an accretion disk 
and plausibly related to unsteady nuclear burning (van der Klis 2006). 
The structure of the flattening in the PSD sample of AXPs and SGRs resembles the VLFN of 
accreting X-ray binaries. 

If one assumes that the variable X-ray flux originates from an accretion  
disk with a size between 1$\times$10$^9$ cm to 1$\times$10$^{12}$ cm, 
(values obtained from Ertan et al. 2007) the range of Keplerian frequencies (r$^3_K$ = GM$_{\rm NS}$/4$\pi^2 \nu_K^2$) that should appear in the continuum noise is then from 0.058 Hz to 2$\times$10$^{-6}$ Hz. This range overlaps with the detected range of broad-band noise in our work which supports that we could be detecting noise from the fall-back accretion disks suggested to exist in these systems (Ertan et al. 2007). However, the variations of this noise does not resemble the case for X-ray binary sources (see Figure 5).  
In X-ray binaries outburst effects last for a few days to a week (where one detects a change in the
broad-band noise level, as well). The persistent level of emission changes with accretion rate
from high states to low states of X-ray binary sources while one also detects 
reallocation of broad-band noise from high frequencies into low frequencies. 
On the other hand, in our entire data set we did not detect any reallocation of noise, the noise we detected always appeared in the 0.05-0.005 Hz range and  our timescale for increased noise  or high level of broad-band noise lasted for over a year or two which is not a detected phenomenon in X-ray binaries. 


\begin{figure}
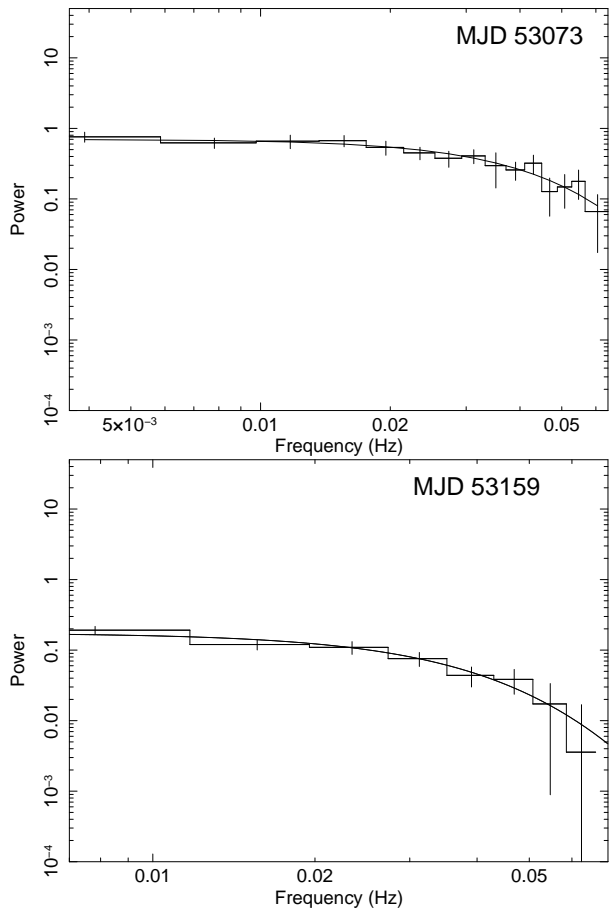

\centerline{
\includegraphics[scale=0.30,angle=270]{1841_90076_3ay_8_256_lb_2_fit.ps}}
\centerline{
\includegraphics[scale=0.30,angle=270]{1708_90076_3ay2_4_128_lb_2_fit.ps}}
\caption{The averaged PSD in time showing the broad-band noise of 1E 1841-045 (top) and RXS J1708-40 (bottom).}
\end{figure}

\begin{figure}
\includegraphics[scale=0.30,angle=270]{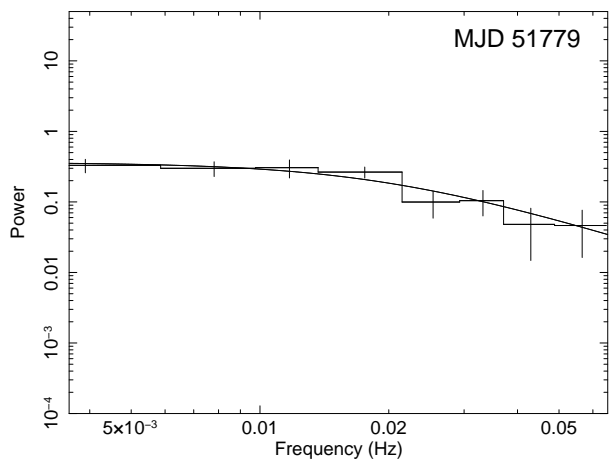} 
\caption{The time averaged PSD of SGR 1806-20 showing the broad-band noise.}
\end{figure}

As discussed in the above paragraphs we favor that the changes in the band
limited low frequency noise in magnetars are related most likely to the burst
characteristics.
Glitches seem to be the least effective of all in the broad-band noise changes from 
these sources. As discussed in the above paragraph, one origin for the detected broad-band noise in AXP and SGRs could be the suggested fall-back disks around these objects with the aforementioned caveats.  

The origin of this noise could also be partially related to the frequency fluctuations of 
the neutron stars as in isolated pulsars and also to the variability that arises from the  
up-scattered radiation in their magnetospheres, particularly enhanced in the presence of a magnetar corona  (Beloborodov \& Thompson 2005, Beloborodov \& Thompson 2007). 
 The plasma coronae in magnetars are a result of the starquakes and twisting of the magnetosphere 
that injects an electric current (i.e., plasma) into the stellar magnetosphere 
(Beloborodov \& Thompson 2005, Beloborodov \& Thompson 2007). 
Emerging currents are maintained during the X-ray outburst and a dense 
thermalized plasma is present in the magnetosphere (lasting from 1-10 years). 
Broad-band noise due to processes such as photon delays between soft and hard X-ray photons are 
well studied in  X-ray binary sources, however such timescales are much shorter 
like a few milliseconds as a result of Compton-up scattering (in the presence of Thompson optical 
depth) from disk coronae (e.g., Vaughan et al. 1994, Schulz \& Wijers
1993) which yields a continuum noise in very high frequencies in comparison with 
the frequency range we detect here (within 256 sec). However, the processes in the
magnetosphere of AXPs and SGRs are more complicated. Lyutikov \& Gavriil (2006) shows that
magnetospheric plasma can distort the thermal X-ray emission emerging from the star surface
by repeated scatterings at the cyclotron resonance (e.g., resonant cyclotron scattering). 

Recently, Rea et al. (2008) showed that an X-ray spectral model that involves resonant 
cyclotron scattering (together with a blackbody model) successfully fits the soft X-ray emission
(1-10 keV data) of ten magnetars. If the dominant scattering process is the resonant scattering
rather than Thompson scattering (as shown by Rea et al. 2008), then since the Thompson optical 
depth to scattering is different than resonant scattering optical depth, the escape times of photons
from the magnetosphere or an existing magnetar corona will be different in comparison with
a simple disk corona where Thompson scattering optical depth dominates (as photons are
Compton up-scattered in the corona). Once escape times differ, so will the time delays
between soft and hard photons will be different in a resonant scattering dominated plasma. 
Rea et al. (2008)  calculates the ratio of optical depths as
$\tau_{res}$/$\tau_T$ = 10$^{5}$ which strongly suggests that given similar geometries
the high level of resonant optical depth to scattering should cause more delays in the
emerging photons. If one assumes that scattering optical depth is directly proportional to
number of scatterings, a millisecond delay then should become about 100 sec delay, that 
could yield noise around 0.01 Hz.
We suggest that especially for 1E 2259+586 this scattering can produce or contribute to the 
band-limited red noise we detect in the 0.005-0.05 Hz band and its changes. 
This is, also, supported by the commensurability of 2 years rise in noise we detect in two AXPs and the
magnetar coronal time scale of 1-10 years (time scale the magnetar corona exists). 
Moreover, the increase in the photon indices of power-law tails in the X-ray spectra: 
For 1E 2259+586, $\Gamma$ increases from -4.3 to -2.8 after the outburst 
(Woods et al. 2004, Kuiper et al. 2006); for 1E 1048.1-5937, $\Gamma$ increases from 
-3.4 to -3.0 slowly after the reported burst at 53185 MJD and this increases to 
-2.3 after the flux outburst in 2007 March (Tiengo et al. 2005; Campana \& Israel 2007, 
Rea et al. 2007, Tam et al. 2008). 

Characteristic of the activity is important for inducing such a process, e.g. a long burst or an outburst like characteristics are necessary to produce such effects of highly increased broad-band noise levels. Glitches or flares show no effect or slight variations in noise levels. This supported by nearly stable band limited noise levels of 1RXS J1708-40 and 1E 1841-045 which show only glitches and no bursts \footnote{http://www.physics.mcgill.ca/$\sim$ pulsar/magnetar/main.html}. Moreover, isolated pulsars that show glitches do not show such long term significant rises in red noise levels in the low frequency band. 

Other models explaining the magnetar timing variabilities are ruled out mostly because of their time 
scales. For example, seismic vibrations that are used to explain QPOs are very rapid and 
connected to giant bursts, and thus is not a satisfactory explanation for broad-band noise. 
One modifying principle for the seismic models is the propagation of Alfv\'en waves towards the core. 
Levin (2006) noted that these oscillations are a continuum, but they decay rapidly thus effectively 
they can not correspond to our persistent rms noise observations. 

\begin{figure*}
\centerline{
\includegraphics[width=9cm,height=13cm,angle=270]{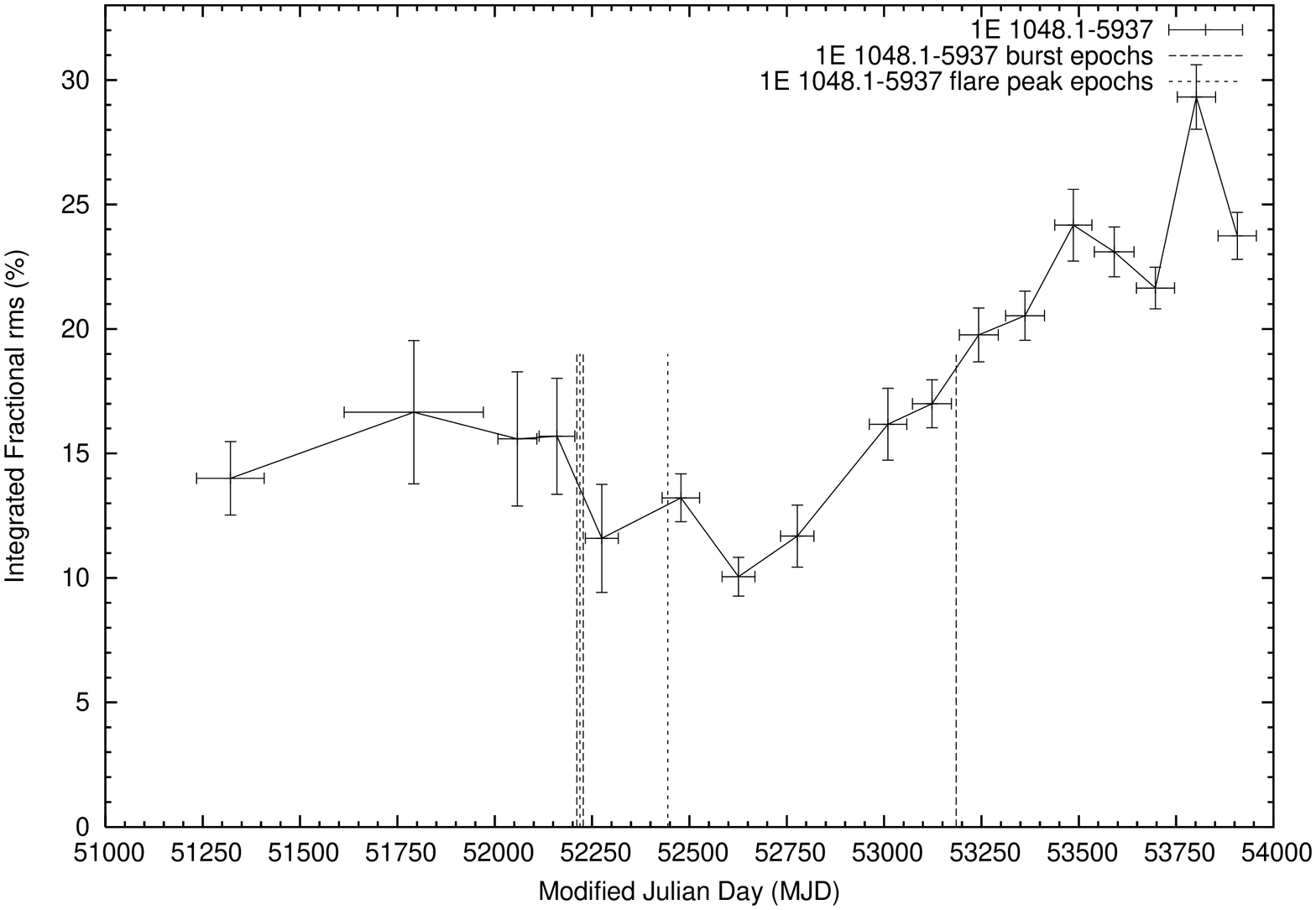}}
\centerline{
\includegraphics[width=9cm,height=13cm,angle=270]{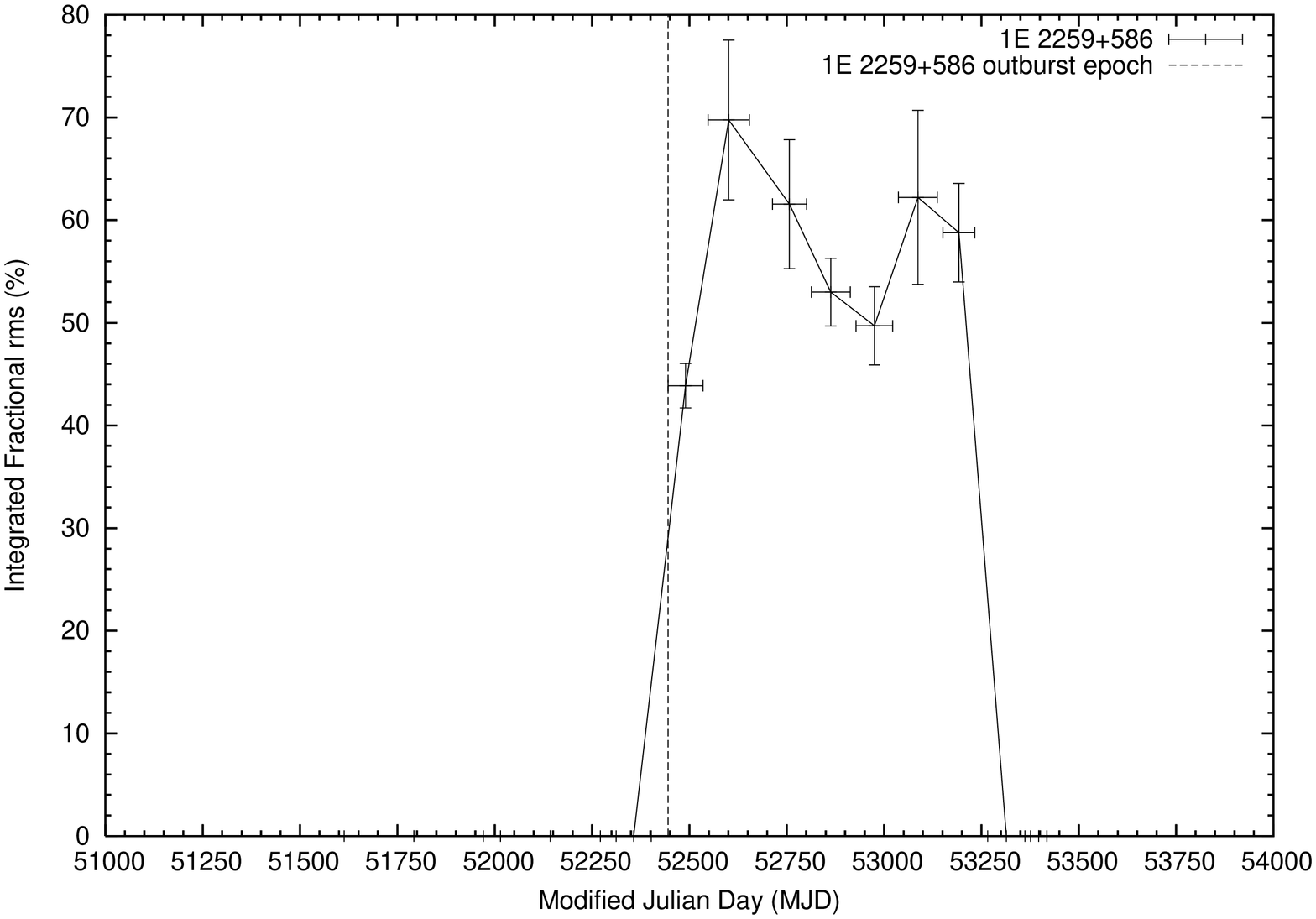}}
\caption{The integrated rms values for two AXPs (1E 2259+586, 1E 1048.1-5937) in time.
The model and the range of integration is described in the text. 
The burst epochs for 1E 1048.1-5937 are from Gavriil, Kaspi \& Woods 2002 and Gavriil, Kaspi \& Woods 2006, flare epochs are from Gavriil \& Kaspi 2004. The outburst epoch for 1E 2259+586 is from Kaspi, 
Gavriil \& Woods 2002.}
\end{figure*}

\section*{Conclusions}

We observed broad-band noise (BLN) in the four AXPs (1E 1048.1-5937, 1RXS 1708-40, 
1E 1841-045 and 1E 2259+586) and an SGR (SGR 1806-20) mainly over the 0.005-0.05 Hz frequency band. 
The integrated rms values were in a range 2.5-15$\%$ for four sources and 0$\%$ rms 
for 1E 2259+586 before bursts. 
After bursts (our data do not include burst/outburst epochs) 20-70$\%$ integrated rms 
noise was observed in two cases; 1E 2259+586 and 1E 1048.1-5937, respectively. 
The  broad-band noise characteristics do not resemble either the accreting X-ray binaries or the 
isolated pulsars completely. 

In general, we found that the four sources indicate a persistent level of integrated rms in 
time over a long base line of  3.5--5.5 years which may be affected by noise from frequency fluctuations, 
 and most likely as a result of variability due to resonant cyclotron scattering in a 
magnetar coronae and/or variability from existing fall-back disks around these objects. 
We also detected large scale changes of the broad-band noise particularly due to 
long burst/outbursts. SGR-like bursts did not produce these type of variations in broad-band red 
noise, however we believe that this is limited with the sensitivity of RXTE (also its background)
and statistical quality of RXTE light curves. 
We calculate that 1E 2259+586 displays high level of broad-band noise lasting for a timescale of 
2 years following its outburst and glitch. In addition, we observed significant rise of the 
rms broad-band noise level in 1E 1048.1-5937 for about 1.95 years, 
after the burst epoch at 53185 MJD. 
We believe that this noise can relate to the existence of remnant disks suggested to 
exist around these objects, but the timescale of changes after burst/outbursts are 
inconsistent with the observations in X-ray binaries. 
We stress that the changes in the broad-band noise levels are related to  
resonant cyclotron up-scattered radiation in AXP and SGR spectra. Correlation with 
long bursts and outbursts indicates an origin in the variability associated with the magnetar coronae. 
The typical timescales of existence of magnetar coronae of 1-10 years is consistent with the 
long term variation of BLN detected in our work.

\section*{Acknowledgments}
We thank an anonymous referee for critical reading of the manuscript which improved this paper greatly.
We, also, thank Altan Baykal and Ali Alpar for helpful comments. BK acknowledges support from T\"UB\.ITAK, The Scientific and Technological Research Council of Turkey,  through project 106T040. SB acknowledges an ESA fellowship and support from the Turkish Academy of Sciences with TUBA-GEBIP (Young Distinguished Scientist award) fellowship. 



\label{lastpage}


\begin{thebibliography}{99}

\bibitem[2001]{alpar01} Alpar M.A., 2001, ApJ., 554, 1245

\bibitem[2000]{b0} Baykal A., Ali Alpar M., Boynton P. E., Deeter J. E., 1999, MNRAS, 306, 207

\bibitem[2000]{b01} Baykal A. , Strohmayer T., Swank J., Alpar M.A., Stark M.J., 2000, MNRAS, 319, 205

\bibitem[2005]{b1} Belloni T., Psaltis D., van der Klis M., 2002, ApJ, 572, 392

\bibitem[2005]{b2} Thompson S., Beloborodov A. M., 2005, ApJ, 634, 565

\bibitem[2007]{b3} Beloborodov A. M., Thompson S., 2007, ApJ, 657, 967

\bibitem[2004]{b4} Borkowski J., Gotz D., Mereghetti S., Mowlavi N., Shaw S., Turler M., 2004, GCN Circ., 2920

\bibitem[2000]{c0} Campana S. M., Israel G. L., 2007, Astron. Telegr., 1043

\bibitem[2000]{c1} Chatterjee P., Hernquist L., Narayan R., 2000, ApJ, 541, 367

\bibitem[1997]{c2} Corcoran M. F., et al., 1997,  Nature, 390, 587

\bibitem[2007]{d2} Dib R., Kaspi V., Gavriil F., 2007a, ApJ, 666, 1152

\bibitem[2007]{d3} Dib R., Kaspi V., Gavriil F., 2007b, preprint (arXiv:0706.4156)

\bibitem[2007]{d4} Dib R., Kaspi V., Gavriil F., Woods P. M., 2007, Astron. Telegr., 1041

\bibitem[2000]{d5} Duncan R.C., Thompson C., 1992, ApJ, 392, L9 

\bibitem[2007]{e1} Ertan, \"U., Alpar, M. A., Erkut, M. H., Ek\c{s}i, K. Y., \c{C}ali\c{s}kan, \c{S}., 2007, Ap\&SS, 308, 73 

\bibitem[2002]{g1} Gavriil F.P., Kaspi V.M., Woods P.M., 2002, Nature, 419

\bibitem[2002]{g2} Gavriil F.P., Kaspi V.M., 2002, ApJ, 567, 1067

\bibitem[2004]{g3} Gavriil F., Kaspi V.M. 2004, ApJ, 607, 959

\bibitem[2004]{g4} Gavriil F. P., Kaspi V. M., Woods Peter M., 2006, ApJ, 641, 418

\bibitem[2001]{h1} Hulleman F., Tennant A. F., van Kerkwijk M. H., Kulkarni S. R., Kouveliotou C., Patel S. K., 2001, ApJ, 563, L49

\bibitem[2001]{h2} Hasinger G., van der Klis M., 1989, A\&A, 225, 79 

\bibitem[2004]{i1} Ibrahim A.I., et al. 2004, ApJ 609,L21

\bibitem[2005]{i2} Israel G.L. et al. 2005, ApJ, 628, L53

\bibitem[1996]{j1} Jahoda K., Swank J. H., Giles A. B., Stark M. J., Strohmayer T., Zhang W., Morgan E. H., 1996, Proc. SPIE, 2808, 59

\bibitem[2000]{k2} Kaspi V. M., Lackey J. R., Chakrabarty D., 2000, ApJ, 537, L31 

\bibitem[2001]{k4} Kaspi V.M., Gavriil F.P., Chakrabarty D., Lackey J.R., Muno M.P., 2001, ApJ, 558, 253

\bibitem[2002]{k1} Kaspi V.M., Gavriil F.P., Woods P. M., 2002, Astron. Telegr., 99

\bibitem[2003]{k3} Kaspi V. M., Gavriil F. P., Woods P. M., Jensen J. B., Roberts M. S. E., Chakrabarty D., 2003, ApJ, 588, L93

\bibitem[2003]{k5} Kaspi V.M., Gavriil F. P., 2003, ApJ, 596, L71

\bibitem[2004]{k6} Kaspi V., Gavriil F., Woods P., Chakrabarty D., 2004, Astron. Telegr., 298

\bibitem[2004]{k7} Kuiper L., Hermsen W., den Hartog P. R., Collmar W., 2006, ApJ, 645, 556

\bibitem[2006]{l1} Levin Y., 2006., MNRS, 368, L35

\bibitem[2006]{l2} Lyutikov M., Gavriil F.P., 2006, MNRAS, 368, 690

\bibitem[1991]{m1} Miyamoto S., Kimura K., Kitamoto S., Dotani T., Ebisawa K., 1991, ApJ, 383, 784

\bibitem[1980]{r1} Ramaty R., Bonazzola S., Cline T. L., Kazanas D., Meszaros P., Lingenfelter R. E., 1980, Nature, 
287, 122 

\bibitem[2007]{r2} Rea N., Tiengo A., Israel G.N., Campana S., 2007, Astron. Telegr., 1121.

\bibitem[2007]{r3} Rea N., Zane S., Turolla R., Lyutikov M., G\"otz D., 2008, ApJ, 686, 1245 

\bibitem[2005]{s1} Strohmayer T. E., Watts A. L., 2005 ApJ, 632, L111

\bibitem[1993]{s2} Schulz N.S., Wijers R.A.M.J., 1993, ApJ, 273, 128

\bibitem[2008]{t2} Tam C. R., Gavriil F. P., Dib R., Kaspi V. M., Woods P. M., Bassa C., 2008, ApJ, 677, 503 

\bibitem[2007]{t3} Tiengo A., Mereghetti S., Turolla R., Zane S., Rea N., Stella L., Israel G.L., 2005, A\&A, 437, 9
97

\bibitem[1995]{t1} Thompson C., Duncan R.C., 1995, MNRAS, 275, 225

\bibitem[2004]{v1} van der Klis M., 2000, ARA\&A, 38, 717

\bibitem[2004]{v2} van der Klis M., 2006, in Lewin W.H.G., van der Klis, M., eds., Cambridge Astron. Ser. Vol. 39,
Compact stellar X-ray sourcesr. Cambridge Univ. Press, Cambridge, preprint (arXiv:0410551) 

\bibitem[1995]{v3} van Paradijs J., Taam R.E., van den Heuvel E.P.J., 1995, A\&A, 299, L41

\bibitem[1994]{v4} Vaughan B.,  van der Klis M., Lewin W.H.G., Wijers R.A.M.J., van Paradijs J., 
Dotani, T., 1994, ApJ, 421, 738 

\bibitem[2007]{w1} Watts, Anna L. \& Strohmayer, Tod E., 2007, AdSpR, 40, 1446

\bibitem[2001]{w2} Wijnands R., 2001, AdSpR, 28, 469

\bibitem[2004]{w3} Woods P.M., Thompson C., 2006, in Lewin W.H.G., van der Klis, M. eds, Cambridge Astron. Ser. Vol. 39, Compact stellar X-ray sources. Cambridge Univ. Press, Cambridge, preprint (arXiv:0406133)

\bibitem[2004]{w4}Woods P.M., Kaspi V.M., Thompson C., Gavriil F.P., Marshall H.L., Chakrabarty D., Flanagan K., Heyl J., Hernquist L., 2004, ApJ, 605, 378


\end{thebibliography}
\end{document}